\documentclass[aps,twocolumn,a4paper,showpacs]{revtex4}
\usepackage[colorlinks=true, pdfstartview=FitV, linkcolor=blue, citecolor=red, urlcolor=magenta,breaklinks=true]{hyperref}
\usepackage{graphicx}
\usepackage[all]{xy}
\usepackage{amsmath}
\usepackage{amssymb}
\usepackage{color}
\usepackage{subeqnarray}
\usepackage{subfigure}
\usepackage{epstopdf}
\usepackage{lmodern} 
\usepackage[T1]{fontenc}
\usepackage{lipsum} 

\newcommand{\bb}{\bibitem}
\newcommand{\bes}{\begin{subequations}}
\newcommand{\ees}{\end{subequations}}
\def\ben{\begin{eqnarray}}
\def\een{\end{eqnarray}}
\newcommand{\bens}{\begin{subeqnarray}}
\newcommand{\eens}{\end{subeqnarray}}

\def\be{\begin{equation}}
\def\ee{\end{equation}}

\begin{document}
\title{Multi-component scalar fields and the complete factorization of its equations of motion} 
\author{Diego R. Granado}
  
\affiliation{Departamento de F\'\i sica Universidade Federal da Para\'iba, 58051-900 Jo\~ao Pessoa PB, Brazil}
\date{\today}

\begin{abstract}

In the paper D.~Bazeia {\textit{et al., EPL}}, {\bf 119}, (2017) 61002, the authors demonstrate the equivalence between the second-order differential equation of motion and a family of first-order differential equations of Bogomolnyi type for the cases of single real and complex scalar field theories with non-canonical dynamics. The goal of this paper is to demonstrate that this equivalence is also valid for a more general classes of real scalar field models. We start the paper by demonstrating the equivalence in a single real scalar model. The first goal is to generalize the equivalence presented in papers by D.~Bazeia {\textit{et al.}} to a single real scalar field model without a specific form for its Lagrangian. The second goal is to use the setup presented in the first demonstration to show that this equivalence can be achieved also in a real multi-component scalar field model, again, without a specific form for its Lagrangian. The main goal of this paper is to show that this equivalence can be achieved in real scalar field scenarios that can be standard, or non-standard, with single, or multi-component, scalar fields.










\end{abstract}
\maketitle


\section{Introduction}
Topological structures are non-perturbative aspects of a theory and can be classified as defect structures. They are static solutions of a classical field theory. In the $\lambda\phi^4$-model, for instance, the defects are static solutions of the equations of motion.\ In \cite{deVega:1976xbp}, the authors show that if classical equations of motion of a model can be reduced to first-order equations this represents a bosonic reduction of a supersymmetric theory.\ This reduction is the core of the first-order framework \cite{FOFGD}.\ These first-order differential equations solutions are defect structures presented in the theory, named as Bogomol'nyi-Prasad-Sommerfield (BPS) states \cite{BPS}.\ The first-order framework deals with systems in which BPS first-order differential equations solutions solve their second-order differential equation of motion.\ Some interesting applications can be found for instance in cosmology.\ Quintessence models are cosmological examples where single real scalar fields can be used to describe the dynamics of the universe \cite{Tsujikawa:2013fta}.\ This model consists of a minimally coupled to gravity canonical real scalar field with a potential.\ It can be used, for instance, to model dark energy. In order to investigate the time evolution of the Hubble parameter in this model the first-order framework was used in \cite{Bazeia:2005tj,Setare:2008ci,Harko:2013gha}.\ Another example of application can be found in \cite{Bazeia:2017mnc}, where the authors use the first-order framework to study kinklike structures in a system with a Dirac-Born-Infeld scalar dynamics. In \cite{CFO1,CFO2}, the authors demonstrate the equivalence between the second-order differential equation of motion and a family of first-order differential equations of Bogomol’nyi type in the bosonic sectors of supersymmetric theories. This result was extended in \cite{Bazeia:2017irs} to the cases of single real and complex scalar fields theories with a non-standard kinetic term. 

As it was pointed in \cite{FOFGD}, the first-order framework can be used also in scenarios with multi-component real scalar fields. Very recently, in order to obtain a better fit with the observational data, a cosmological model called multiquintessence was suggested in \cite{Correa:2016cgj}. This model uses multi-component real scalar fields and the authors use the first-order framework to obtain the scalar fields solutions. Besides cosmology, the first-order framework with multi-component scalar fields was also used, for instance, in braneworld models \cite{FIRST3,Dutra:2014xla,Xie:2019jkq,Farokhtabar:2016fhm}. The main goal of this paper is to extend the demonstration found in \cite{CFO1,CFO2,Bazeia:2017irs} to a system described by multi-component real scalar fields.

This paper is organized as follows: in the section \ref{sec-1} we extend the demonstration presented in \cite{CFO1,CFO2,Bazeia:2017irs} for the single real scalar field case to a case where the Lagrangian can develop a more general dependence on the kinetic and potential terms; in the section \ref{sec-2}, we treat the multi-component scalar field case. We work in $(1+1)$-dimensional spacetime with metric $(+,-)$ and for simplicity we adopt dimensionless field, spacetime coordinates and coupling constants.

\section{SINGLE SCALAR FIELD} \label{sec-1}
In this section we generalize the setup used in the Refs. \cite{Bazeia:2017irs,CFO1,CFO2}. In \cite{CFO1}, the authors demonstrate the equivalence between the equation of motion and a family of BPS first-order differential equations for the case of a real single scalar field with a canonical dynamics. In \cite{Bazeia:2017irs}, this result was extended to a single real scalar field with non-canonical dynamics. In both works the authors demonstrate the equivalence by considering an specific form for the Lagrangian in each work. In this section we will follow the generalities presented in \cite{FOFGD} by considering only the Lagrangian, ${\cal L}\left(\phi,X\right)$, without writing a specific form for it. 

The most general action for a single real scalar field reads
\be
\label{lagran0}
S=\int d^2x {\cal L}\left(\phi,X\right)
\ee
where $X$ reads
\be
X=\frac{1}{2}\partial_\mu\phi\partial^\mu\phi
\ee
The energy-momentum reads
\be
T_{\mu\nu}={\cal{L}}_{X}\partial_\mu\phi\partial_\nu\phi-g_{\mu\nu}{\cal{L}}
\ee
where $d{\cal{L}}/dX={\cal{L}}_{X}$. We are working with the static field configurations $\phi(t,x)=\phi(x)$. Thus we can write the energy-momentum component in the form:
\begin{eqnarray}
\rho(x)&=&-{\cal{L}}\\
\tau(x)&=&{\cal{L}}_{X}\phi'\phi'+{\cal{L}}
\label{stress1}
\end{eqnarray}
The equation of motion from \eqref{lagran0} reads:
\be
\partial_\mu({\cal{L}}_{X}\partial^\mu\phi)={\cal{L}}_{\phi}
\label{eom0}
\ee
The equation of motion can be recast in the following form
\be
G^{\alpha\beta}\partial_\alpha\partial_\beta\phi=-2X{\cal{L}}_{X}\phi+{\cal{L}}_{\phi}
\ee
where 
\begin{eqnarray}
G^{\alpha\beta}&=&{\cal{L}}_{X}\eta^{\alpha\beta}+{\cal{L}}_{XX}\partial^\alpha\phi\partial^\beta\phi
\end{eqnarray}
In the static configuration we have:
\be
\left({\cal{L}}_{X}+2{\cal{L}}_{XX}X\right)\phi''=2X{\cal{L}}_{X\phi}-{\cal{L}}_{\phi}
\ee
As it was shown in the Ref. \cite{FOFGD}, the equation above can be integrated to give:
\be
{\cal{L}}-2{\cal{L}}_{X}X=C
\label{integrationcte}
\ee
where $C$ is an integration constant that can be indentified with the energy-momentum component \eqref{stress1} \cite{FOFGD}. A stressless condition is demanded from stability \cite{SPGD}. Thus we have
\be
{\cal{L}}-2{\cal{L}}_{X}X=0
\label{stability}
\ee
From this constraint the energy density can be written as
\be
\rho(x)=-{\cal{L}}={\cal{L}}_{X}\phi'\phi'
\ee
Following the formalism presented in \cite{FOFGD}, from the equations of motion \eqref{eom0} we can define a function of the scalar field $W=W(\phi)$ as
\be
{\cal{L}}_{X}\phi'=W_{\phi}
\label{w0}
\ee
and the energy density can be written as
\be
\rho(x)=W_{\phi}\phi'=\frac{d W}{dx}
\ee
Therefore the energy reads
\begin{eqnarray}
E=\Delta W&=&W(\phi(\infty))-W(\phi(-\infty))
\end{eqnarray}
Using eq. \eqref{w0} the equation of motion \eqref{eom0} reads
\be
W_{{\phi}{\phi}}\phi'=-{\cal{L}}_{\phi}
\label{lphi}
\ee
The equation \eqref{lphi} is a fisrt-order diferential equation, which together with the eq. \eqref{w0} provides solutions for the equation of motion. 

In order to achieve the main result of this works, \textit{i.e.}, that we can find BPS solutions that are equivalent to the solutions of the equation of motion, as in Refs.\cite{Bazeia:2017irs,CFO1, CFO2}, we define a quantity $R(\phi)$ as
\be
R(\phi)=\frac{{\cal{L}}_{X}\phi'}{W_{\phi}}
\label{R}
\ee 
From that we have
\begin{eqnarray}
\label{20}
\frac{d R(\phi)}{dx}&=&\left(W_{\phi}\frac{d}{dx} ({\cal{L}}_{X}\phi')-{\cal{L}}_{X}\phi'^2W_{\phi\phi}\right)\frac{1}{W_{\phi}^2}\nonumber\\
&=&\Big[W_{\phi}-{\cal{L}}_{X}\phi'\Big]\frac{(W_{{\phi}{\phi}}\phi')}{W_{\phi}^2}\nonumber\\
\end{eqnarray}
where we have used the second-order equation of motion for static configurations, $ ({\cal{L}}_{X}\phi')'=-{\cal{L}}_{\phi}$ and the relation: ${\cal{L}}_{X} {\cal{L}}_{\phi}=W_{\phi}W_{\phi\phi}$.
Turning our attention to the following quantity $S(\phi)=W_{\phi}-{\cal{L}}_{X}\phi'$,  one can see that 
\ben
\frac{d S(\phi)}{dx}&=&W_{\phi\phi}\phi'-\frac{d}{dx} ({\cal{L}}_{X}\phi') \nonumber\\
&=& W_{\phi\phi}\phi'-\left(\phi'{W_{\phi\phi}}\right) \nonumber \\
&=&0
\label{ds}
\een
Thereore the quantity $S(\phi)$ does not vary with $x$ when $\phi(x)$ is the solution of the equation of motion. 

In the introduction we have mentioned that the $\lambda\phi^4$-model supports BPS topological solutions.\ This occurs because of the $\lambda\phi^4$-potential. This means that in order to describe the spontaneous symmetry breaking, and to reveal the topological structures, boundary conditions over the scalar fields are necessary to minimize the potential and to break the global symmetry. As used in \cite{CFO2}, the boundary conditions that support BPS topological solution in the $\lambda\phi^4$-potential are: $\lim_{x\rightarrow - \infty} \phi(x)= v_{k}$, where $v_{k}$ are the vacuum states, and $\lim_{x\rightarrow - \infty} \phi'(x)= 0$. Here, we will use these same boundary conditions and, at the same time, assume that the vacuum states are extrema of the function $W(\phi)$, {\it{i.e}}.,  $\lim_{x\rightarrow - \infty} W_{\phi}= 0$. Thus we have
\be
\lim_{x\rightarrow - \infty} S(\phi)= \lim_{x\rightarrow - \infty} \left(W_{\phi}-({\cal{L}}_{X}\phi')\right)=0.
\label{limds}
\ee
Therefore, we  have that $S(\phi)$ vanishes. From this we get $R=\pm 1$ which provides: ${\cal{L}}_{X}\phi'= W_{\phi}$.  This result shows the equivalence between the solution of the equations of motion, $ ({\cal{L}}_{X}\phi')'=-{\cal{L}}_{\phi}$, and the BPS solutions ${\cal{L}}_{X}\phi'= W_{\phi}$. This result is a extension of what was introduced in Refs. \cite{Bazeia:2017irs,CFO1, CFO2} to the case of a single real scalar with the general first-order framework introduced in Ref. \cite{FOFGD}.

One important remark should be pointed out before we move to the next section. As it was pointed in \cite{CFO1}, the result ${\cal{L}}_{X}\phi'= W_{\phi}$ could be obtained in the equation \eqref{20} without defining the quantity $S(\phi)$. We preferred to define $S(\phi)$ along the computation only to setup a framework to be used in the next section.

\section{MULTI-COMPONENT SCALAR FIELDS} \label{sec-2}
In this section we extend the results presented in the previous section by treating the multi-component real scalar field case.\ We keep the same dimensionality and conventions adopted before.

The most general multi-component real scalar filed action reads
\be
\label{lagran01}
S=\int d^2x {\cal L}\left(\phi_i,X_{jk}\right)
\ee
where $\phi_i$ is a set of $n$ scalar fields $\{\phi_1,\phi_2,\dots,\phi_n\}$ and $X_{jk}$ reads
\be
X_{jk}=\frac{1}{2}\partial_\mu\phi_j\partial^\mu\phi_k
\ee
In this case the energy-momentum 
components read
\begin{eqnarray}
\rho(x)&=&-{\cal{L}}\\
\tau(x)&=&{\cal{L}}_{X_{ij}}\phi_i'\phi_j'+{\cal{L}}
\label{stress}
\end{eqnarray}
From the Lagrangian \eqref{lagran01} we have the following equation of motion
\be
\partial_\mu({\cal{L}}_{X_{ij}}\partial^\mu\phi_j)={\cal{L}}_{\phi_i}
\label{eom}
\ee
Following the steps of the previous section the equation of motion can be written as:
\be
\left({\cal{L}}_{X_{ij}}+2{\cal{L}}_{X_{ij}X_{jm}}X_{lm}\right)\phi_j''=2X_{jl}{\cal{L}}_{X_{ij}\phi_l}-{\cal{L}}_{\phi_i}
\label{integrationctemulti}
\ee
As discussed in the previous section in the eq. \eqref{integrationcte} and as it was pointed out in Ref. \cite{FOFGD}, eq. \eqref{integrationctemulti} can be integrated to give a constant that can be identified with the pressure in eq. \eqref{stress}. As in eq. \eqref{stability}, stability  demands the following condition \cite{SPGD}:
\be
{\cal{L}}-2{\cal{L}}_{X_{ij}}X_{ij}=0
\ee
Thus the energy density can be written as
\be
\rho(x)=-{\cal{L}}={\cal{L}}_{X_{ij}}\phi_i'\phi_j'
\label{energydensity}
\ee
As in eq. \eqref{w}, we can define a function $W=W(\phi_1,\phi_2,\dots,\phi_n)$ such that
\be
{\cal{L}}_{X_{ij}}\phi_j'=W_{\phi_i}
\label{w}
\ee
and rewrite eq. \eqref{energydensity} as
\be
\rho(x)=W_{\phi_i}\phi_i'=\frac{d}{dx}W
\ee
and the energy for this case reads
\begin{eqnarray}
E=\Delta W&=&W(\phi_1(\infty),\phi_2(\infty),\dots,\phi_n(\infty))\nonumber\\
&-&W(\phi_1(-\infty),\phi_2(-\infty),\dots,\phi_n(-\infty))~~
\end{eqnarray}
From eq. \eqref{w}, we can rewrite eq. \eqref{eom} as
\be
W_{{\phi_{i}}{\phi_{j}}}\phi_j'=-{\cal{L}}_{\phi_i}
\label{33}
\ee
In this case, the analog of the scalar quantity definied in eq. \eqref{R}, reads:  
\be
R(\phi_i)=\frac{c_i{\cal{L}}_{X_{ij}}\phi_j'}{c_lW_{\phi_l}}
\ee 
where $c_i$ is a constant unitary vector. Thus we have
\begin{eqnarray}
\frac{d}{dx}\left(R(\phi_i)\right)&=&\frac{c_i}{c_lW_{\phi_l}}\frac{d}{dx}\left({{\cal{L}}_{X_{ij}}\phi_j'}\right)+ {c_i{\cal{L}}_{X_{ij}}\phi_j'}\frac{d}{dx}({c_lW_{\phi_l}})^{-1}\nonumber\\
&=&\Big[{c_lW_{\phi_l}}-{c_i{\cal{L}}_{X_{ij}}\phi_j'}\Big]\frac{({c_i}W_{{\phi_{i}}{\phi_{j}}}\phi_j')}{({c_lW_{\phi_l}})^{2}}\nonumber\\
\end{eqnarray}
where we have used the equation of motion, $({{\cal{L}}_{X_{ij}}\phi_j'})'=-{\cal{L}}_{\phi_i}$, and the equation \eqref{33}.\ Defining a quantity $S(\phi_i)={c_lW_{\phi_l}}-{c_i{\cal{L}}_{X_{ij}}\phi_j'}$ we can see that 
\ben
\frac{d S(\phi_i)}{dx}&=&c_lW_{\phi_l\phi_k}\phi_k'-\frac{d}{dx} ({c_i{\cal{L}}_{X_{ij}}\phi_j'}) \nonumber\\
&=& c_lW_{\phi_l\phi_k}\phi_k'-\left(c_iW_{\phi_i\phi_n}\phi_n'\right) \nonumber \\
&=&0, \nonumber
\een
This is the same result obtained in \eqref{ds}\@.\ {Following} the prescription introduced in the previous section, in order to have BPS topological solutions we impose: $\lim_{x\rightarrow - \infty} \phi_k(x)= v_{k}$, where $v_{k}$ are the vacuum states, $\lim_{x\rightarrow - \infty} \phi_k'(x)= 0$ and we assume that the vacuum states are extrema of the function $W(\phi)$, {\it{i.e}}.,  $\lim_{x\rightarrow - \infty} W_{\phi}= 0$. As in \eqref{limds} we obtain
\be
\lim_{x\rightarrow - \infty} S(\phi)= \lim_{x\rightarrow - \infty} \left({c_lW_{\phi_l}}-{c_i{\cal{L}}_{X_{ij}}\phi_j'}\right)=0.
\ee
As it can be seen this is the same result obtained in the previous section: $R=\pm 1$ then ${\cal{L}}_{X_{ij}}\phi_j'= W_{\phi_i}$. As it was stated before this result means the equivalence between the solutions of the equation of motion, $({{\cal{L}}_{X_{ij}}\phi_j'})'=-{\cal{L}}_{\phi_i}$ and the BPS solution ${\cal{L}}_{X_{ij}}\phi_j'= W_{\phi_i}$.

\section{Comments and conclusions}\label{sec-com}
In the light of the general setup proposed in the Ref. \cite{FOFGD}, the present work investigates the equivalence between the equation of motion and a BPS solution in single and multi-component real scalar field models in $(1 + 1)$-dimensions.

In the Ref.~\cite{CFO1,CFO2}, the authors show the equivalence between the equations of motion and BPS first-order equation in the Wess-Zumino model. In the Ref. \cite{Bazeia:2017irs}, the authors show that this demonstration is valid in a scalar field model with non-canonical dynamics. In our work we demonstrate that this equivalence is valid for the most general scalar field model by treating the single and multi-component scalar field cases. Our computations support the idea that the equivalence still holds in stantard, or non-standard, scenarios with single, or multi-component, real scalar fields.

 As mentioned in the introduction, multi-component real scalar fields scenarios are being used in cosmology to create models more compatible with the recent data. Due to the generality of our model and the current interest in multi-component scalar field scenarios, we believe that our result and computations can be relevant for the development of more general scalar field models in cosmology and in high energy physics.

\section*{Acknowledgments}
The author thanks Dionisio Bazeia for the discussions and CNPq  for the financial support. 

\end{document}